\documentclass[12pt]{article}

\usepackage{amsmath}
\usepackage{amssymb}
\usepackage{graphicx}
\usepackage{adjustbox}
\usepackage{listings}
\usepackage{multirow}
\usepackage{multicol}
\usepackage{amsthm}
\usepackage{algorithm}
\usepackage{algorithmic}
\usepackage{url}
\usepackage{bm}
\usepackage{array}
\usepackage{wrapfig}
\usepackage{float}

\usepackage{epstopdf}
\usepackage{subfigure}

\usepackage[numbers,sort&compress]{natbib}
\bibpunct[, ]{[}{]}{,}{n}{,}{,}

\theoremstyle{plain}
\newtheorem{theorem}{Theorem}[section]

\theoremstyle{definition}
\newtheorem{definition}[theorem]{Definition}

\theoremstyle{remark}

\begin{document}


\title{Modulation Graphs in Popular Music}

\author{Jason I. Brown\footnote{jason.brown@dal.ca, corresponding author} 
 ~and 
Ian George  \\ Department of Mathematics and Statistics, Dalhousie University\\ \\}

\maketitle

\begin{abstract}
In this paper, graph theory is used to explore the musical notion of tonal modulation, in theory and application.  We define (pivot) modulation graphs based on the common scales used in popular music.  Properties and parameters of these graphs are discussed. We also investigate modulation graphs for the canon of Lennon-McCartney songs in the works of The Beatles.  Our approach may provide composers with mathematical insights into pivot modulation.

\end{abstract}

\section{Introduction}

Graph theory offers a unique perspective into the sequential nature of music.  For example, Crans \cite{crans} explores the actions of the dihedral group on the set of major and minor triads which are elegantly represented on the triangular lattice known as the Neo-Riemannian Tonnetz \cite{riemann}, on Douthett and Steinbach's hexagonal lattice \cite{douthett}, and on Waller's torus \cite{waller}.  Graphs also serve as a way to study voice leading \cite{wixey}.  

In addition, graphs can be used to model the possibilities of tonal modulation; it is this last idea which is of special interest to us. Tonal modulation is the act of changing the key in a composition, and can be central to the emotional portrait of a piece of music.  Examples of modulation can be found in ``Here, There and Everywhere'' by The Beatles (G major $\rightarrow$ B$\flat$ major $\rightarrow$ Gmajor) and ``Money, Money, Money'' by ABBA (A minor $\rightarrow$ B$\flat$ minor).  Graphs can naturally constructed with the vertices being keys, and (directed) edges connecting the vertices where either modulations are possible, or present in a database of songs.
Walton \cite{walton} introduced two such graphs, one based on major scales, the other on both major and natural minor scales, and restricted modulation to that which is \emph{simple}, in which one changes keys via a shared (or ``pivot'') major or minor chord.  We consider here a broader range of {\em pivot modulation} graphs, based on an extended set of scales (and modes). We consider the combinatorial properties of such graphs, and perform an analysis of modulations present in the discography of The Beatles to bridge our theoretical model to what is done in practice.  

Ultimately, it is a composer's discretion as to what modulations take place and how in a composition.  Our aim is to set modulations in a mathematical setting, with the expectation that a combinatorial view might provide new musical insights and options.

\section{Musical and Mathematical Background}

We assume the reader is acquainted with the basics of western music theory and graph theory (see, for example, \cite{west} for the latter). 
We introduce the numeric labelling of the octave, $C \leftrightarrow 0, D\flat = C\sharp \leftrightarrow 1 , ..., B \leftrightarrow 11$, a bijection between the pitch classes (which we often refer to as {\em notes}) and $\mathbb{Z}_{12}$; addition of pitch classes corresponds to addition modulo 12.  
To create an interesting musical landscape, a sequence of notes can be chosen to form a {\em scale}, the type of which is invariant under addition modulo 12.  
Taking all 12 notes corresponds to the {\em chromatic scale}, but the most common scales -- such as the family of diatonic scales (i.e. major, minor, and their modes -- contain 7 notes.  The initial note of a scale is called the {\em tonic}, providing a sense of stability or ``home'' for the scale.  The scale that forms the essential foundation of a song (or part of a song) is called the {\em key} (while there are algorithms to determine the key of a song, an accurate account of the key of a song often requires musical analysis of the melody and chords). 

There are five scales, important in pop and rock music, that are worthwhile to mention for later discussion of modulation.  
\begin{itemize}
\item The first, and perhaps the most fundamental to western music, is the {\em major scale}: 
if $i$ is taken to be the tonic, then the major scale is formed as: 
\[ i, i+2, i+4, i+5, i+7, i+9, i+11\] where addition, as throughout this paper, is taken modulo 12.  For example, the $C$ major scale has root $C$ and contains the notes $C, D, E, F, G, A, B$.  
\item Next, we have the {\em mixolydian scale}: starting from tonic $i$, the scale contains \[i, i+2, i+4, i+5, i+7, i+9, i+10;\] for example, the $C$ mixolydian scale is $C, D, E, F, G, A, B\flat$.  The mixolydian scale differs from the major scale only by a flattened seventh note. One can view this scale as arising from a major scale with tonic seven semitones down by starting on the fifth note of the major scale (viewed in this way, the mixolydian scale is often called a {\em mode} of the corresponding major scale). The mixolydian scale plays a prominent role in rock music.
\item Another scale which is a mode of the major scale is the {\em natural minor scale}.  This scale is the mode built on the sixth note of a major scale, and is known as the aeolian mode.  The natural minor scale with tonic $i$ consists of the notes \[i, i+2, i+3, i+5, i+7, i+8, i+10.\]  
The $C$ natural minor scale is $C, D, E\flat, F, G, A\flat, B\flat$.  The {\em relative minor scale} of a major scale with tonic $i$ has tonic $i-3$ and shares exactly the same notes as its {\em relative major scale}.  
\item A variation of the natural minor scale is the {\em harmonic minor scale}, with a sharpened seventh note: starting on tonic $i$, the scale consists of the notes \[i, i+2, i+3, i+5, i+7, i+8, i+11.\]  Hence, the $C$ harmonic minor scale would be $C, D, E\flat, F, G, A\flat, B$.  The harmonic minor scale is prominent in music originating in eastern Europe, but has permeated popular music as well. 
\item Lastly, we mention another variation of the natural minor, called the {\em melodic minor}.  This scale has sharpened sixth and seventh notes, so alternatively this scale can be thought of as a major scale with its third note flattened.  The form of the melodic minor is \[i, i+2, i+3, i+5, i+7, i+9, i+11,\] with the $C$ melodic minor being $C, D, E\flat, F, G, A, B$. (In some settings, the melodic minor in descending form contains the notes of the natural minor scale with the same tonic). Melodic minor scales appear often in jazz settings, but also may appear in pop music as well.
\end{itemize}

Scales provide a way to build chords with three notes ({\em triads}).  Suppose our scale is the sequence $j_0,j_1,\ldots,j_6$. then for any $k \in \{0,1,\ldots,6\}$, the set $\{j_k,j_{k+2},j_{k+4}\}$ (addition modulo $7$ here) is a ({\em diatonic}) triad (or simply a chord in our context) of the scale. First, choose a note in a scale to be the root of the triad.  
In this way, every note of a scale has a corresponding diatonic triad built upon it, and a note of the scale appears in three such triads.  Depending on the context, diatonic triads may be major (i.e. of the form $\{i,i+4,i+7\}$), minor (of the form $\{i,i+3,i+7\}$), diminished (of the form $\{i,i+3,i+6\}$), or augmented (of the form $\{i,i+4,i+8\}$).
(It is also important to make a remark about {\em transposition}.  Transposition is the application of a function $f_k:x \rightarrow x+k$ (addition of course modulo 12) for some fixed $k \in {\mathbb Z}_{12}$. Transposition preserves the nature of both scales and chords.)

What does it mean to modulate in a piece of music?  A {\em modulation} is a change of key at some moment in a composition.  
According to the {\em Berklee Book of Jazz Harmony} \cite{berklee}, there are a number of ways to achieve a modulation.  {\em Direct modulation} involves reaching a new key abruptly, with no transition. In a {\em pivot modulation}, a chord common to an original and target key is used in a progression to reach the target.  A third kind of modulation, called {\em transitional}, involves repeating harmonic progressions (i.e. sequences) that bridge two keys in a key-less fashion.  

Our focus in this paper will be on pivot modulation.
For some mathematical clarity, let us introduce the following notation: for a scale $S$, let $T(S)$ denote the set of diatonic triads from $S$.

\begin{definition}
Given two scales $S$ and $P$, we say there exists a (pivot) modulation between $S$ and $P$ (or between $P$ and $S$) if $T(S) \cap T(P) \neq \emptyset$.  If a modulation exists, we call every $p \in T(S) \cap T(P)$ a pivot chord (or pivot triad) of $S$ and $P$.
\end{definition}

\noindent This treatment of modulation is very similar to that from \cite{walton}, except here we permit diminished and augmented triads to be pivot chords. While these types of diatonic chords do not play a role when one restricts to pivots between major scales or natural minor scales (as each such scale contains no augmented chord and a unique diminished chord that is shared with no other except the relative major/minor), such chords will play a role in our discussion of our extended scale collection. 

\section{Pivot Modulation Graphs}

From our definition of the existence of a pivot modulation above, we can now give a construction of a simple graph of pivot modulation: a collection of scales $\Sigma$ is taken to be the vertex set, with an edge $e = \{S_1,S_2\}$ if and only if $S_1$ and $S_2$ are distinct scales in $\Sigma$ with $T(S_1) \cap T(S_2) \neq \emptyset$, that is, if and only if $S_1$ and $S_2$  share a (pivot) chord.  
If $E$ is the set of all such edges, then the pivot graph $G = (\Sigma, E)$ is a {\em simple} graph, that is, without multiple edges.   
Alternatively, we can form a multigraph in a similar fashion, where we allow $|T(S_1) \cap T(S_2)|$ edges between vertices $S_1$ and $S_2$.  Each of these edges can be labelled with a unique element from $T(S_1) \cap T(S_2)$.  Such a multigraph would be useful for examining sequences of pivot modulations as walks in the correspondiing multigraph.  We may also wish to ``combine'' several scales so that they are represented by a single vertex.  To do so, we simply take the union of triad sets for each scale: $T(F) = \bigcup_{S \in F} T(S)$, for a collection of scales $F$.

Now that we have shown how to form a graph of pivot modulation, how might they be of use or interest?  First and foremost, they are a novel tool for visualizing modulation.  We may want to know some combinatorial properties of these graphs such as diameter and clique number, or more physical properties including cliques, independent sets, and walks.  From a purely mathematical position such information is intriguing, but it can also be useful for a musician or composer.  For example, suppose that a composer wants to pass from a certain key to another, passing through two intermediate keys.  A sequence of pivot modulations with the desired effect may be found by examining the walks of length 3 between the initial and target keys of a pivot modulation graph.  If a composer would like to select a number of keys between any two of which there is a modulation, the cliques of a graph would give such groupings.  In this section we aim to discuss graphs which represent a realistic framework for pivot modulation in pop music and some of their properties. We remark that our approach here extends that of \cite{walton}, where only pivot modulation graphs related to major scales and natural minor scales were considered (we do not revisit those here). Our extension was motivated by the fact that some of the best pop songs involve other scales, and that mathematical tools for modulation in those settings would be very useful. 

\subsection{Major/Mixolydian Scales}

\begin{figure}[h]
    \centering
    \includegraphics[scale=0.5]{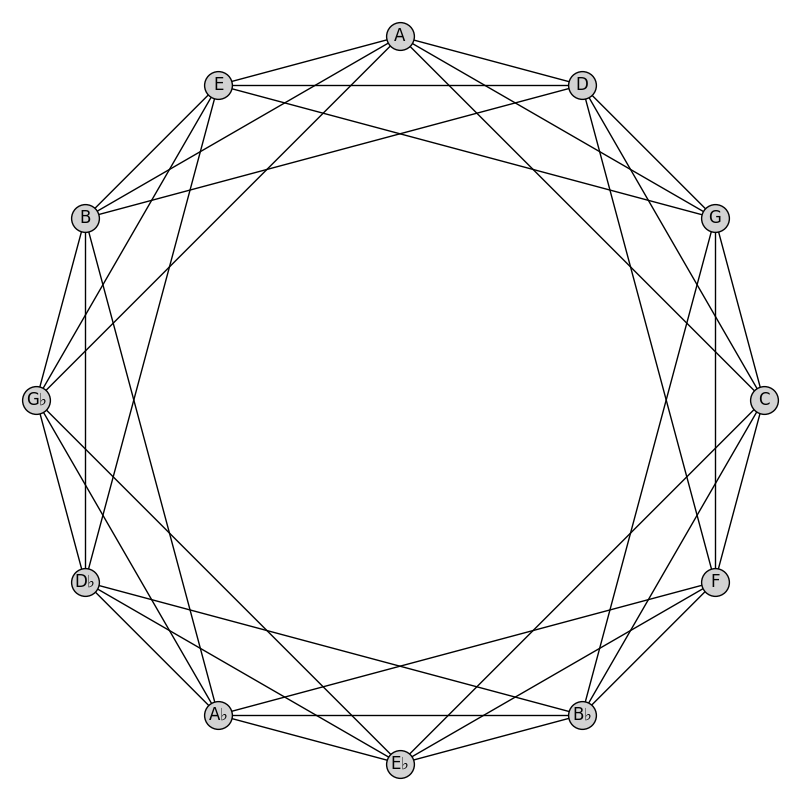}
    \caption{The 12-vertex pivot modulation graph where each vertex represents the major and mixolydian scales sharing a tonic.  This graph can be described as the circulant graph $C_{12}(2, 3, 5)$ with distances measured in semitones.}
    \label{fig:major_graph}
\end{figure}

The first pivot modulation graph we would like to consider is the simple graph formed from major scales and their mixolydian modes (see Figure~\ref{fig:major_graph}).  In this graph, a given vertex represents a major scale and the corresponding mixolydian mode (i.e. sharing the same tonic).  For example, the $C$ vertex represents the $C$ major scale and the $C$ mixolydian scale.  We may notate such a pairing with a capital ``$M$'', for major, subscripted with the numeric label of the relevant tonic: $M_i$.  Our decision to combine these two scales into a single vertex comes from their compatibility: as noted earlier, the major and mixolydian scales only differ by a single note.  
In several genres of modern music, such as rock and roll or pop, borrowing chords from each scale to create a hybrid musical landscape of major/mixolydian is especially common.  
Thus we have 12 vertices in our graph that we will denote by their tonics.

This pivot modulation graph is in fact the circulant graph $C_{12}(2, 3, 5)$ (a {\em circulant graph} $C(n,S)$ has vertex set ${\mathbb Z}_{n}$ with $S \subseteq {\mathbb Z}_{n}$ and edge set $\{\{i,i+s\}: s \in S\}$).  Observe that the arrangement of vertices follows the \emph{circle of fifths}.
This graph has diameter 2, independence number 3 and clique number 4.  Thus to modulate from a given major/mixolydian scale to any other, at most 1 intermediate modulation is required, the largest number of scales of which any pair does not have a modulation is 3, and there are at most 4 scales where all pairs have a modulation between them.

Let us now examine the maximal independent sets and cliques of this graph.  
There are two classes of maximal independent set with 2 vertices: the first contains vertices a semitone apart, and the other a tritone apart.  The only {\em maximum} independent sets (that is, those of maximum cardinality) are formed from scales whose tonics form augmented triads, thus having 4 in total.  All of the maximal cliques are also maximum (and of cardinality $4$) and of one class, having the form of a sequence of four scales whose tonics are consecutively separated by a perfect $5^{th}$.

\begin{table}[h]
    \centering
    \begin{tabular}{|c||c|}
        \hline
        Class of Set & Count \\
        \hline\hline
        $\{ M_i, M_{i+1} \}$ & 12 \\
        $\{ M_i, M_{i+6} \}$ & 6 \\
        $\{ M_i, M_{i+4}, M_{i+8} \}$ & 4 \\
        \hline
    \end{tabular}
    \vspace{2mm}
    \caption{Maximal independent sets of the 12-vertex major/mixolydian pivot modulation graph.}
    \label{tab:maj_ind_set}
\end{table}

\begin{table}[h]
    \centering
    \begin{tabular}{|c||c|}
        \hline
        Class of Set & Count \\
        \hline\hline
        $\{ M_i, M_{i+2}, M_{i+7}, M_{i+9} \}$ & 12 \\
        \hline
    \end{tabular}
    \vspace{2mm}
    \caption{Maximal cliques of the 12-vertex major/mixolydian pivot modulation graph.}
    \label{tab:maj_clique}
\end{table}

As for the automorphisms of the graph, they are described by the following group generators: $(M_0, M_1, M_2, ..., M_{11}), (M_1, M_{11})(M_2, M_{10})(M_3, M_9)(M_4, M_8)(M_5, M_7)$.  This group is isomorphic to $D_{12}$, the dihedral group on 12 elements.

\subsection{Minor Scales}

\begin{figure}[h]
    \centering
    \includegraphics[scale=0.5]{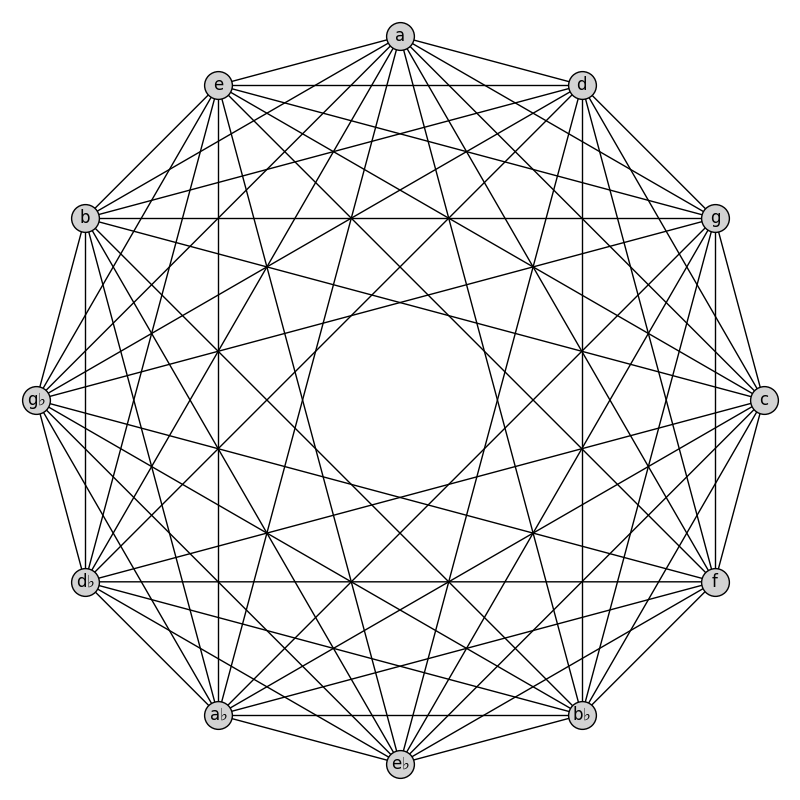}
    \caption{The 12-vertex pivot modulation graph where each vertex represents the natural, harmonic and melodic minor scales sharing a tonic.  This graph can be described as the circulant graph $C_{12}(1, 2, 3, 4, 5)$, or the complete graph $K_{12}$ without a perfect matching.}
    \label{fig:minor_graph}
\end{figure}

Now we turn our attention to minor scales -- the natural, harmonic, and melodic minor.  The graph we consider in this section is again one with 12 vertices (see Figure~\ref{fig:minor_graph}).  Now, however, its vertices will be the union of diatonic triad sets from the natural, harmonic, and melodic minor scales sharing a tonic.  Our justification for such a choice is in the same spirit as before: these flavours of minor scales may be easily interwoven.  According to \cite{berklee}, ``Minor key jazz tunes are rarely diatonic to just one of these scales.  Typically, chords from several of these sources are used interchangeably in minor key compositions.''  Thus to keep to a framework that is realistic for pivot modulation, we will adopt the practice of grouping each minor scale into a single vertex.  This grouping of scales, similar to above, can be denoted as $m_{i}$ where $i$ is the tonic of each minor scale.

This pivot modulation graph can be simply described as the circulant graph $C_{12}(1, 2, 3, 4, 5)$, that is, the complete graph $K_{12}$ minus a perfect matching 
\[ \{\{m_0, m_6\}, \{m_1, m_7\}, \{m_2, m_8\}, \{m_3, m_9\}, \{m_4, m_{10}\}, \{m_5, m_{11}\} \}\] -- the set of edges joining vertices separated by a tritone (that is, by six semitones).  Thus one can modulate, in a single step, from a minor scale grouping to any other except to the scale a tritone away.  Hence it is clear that this graph has diameter 2, which is equal to its independence number.  In contrast, it has clique number 6.

This highly connected graph only permits a single class of maximal independent set, that of the tritone-paired vertices.  However, there are a number of classes of maximal cliques.  Each of the 6 classes of maximal cliques are of maximum size, with 2 notable classes being: vertices forming a semitone sequence (i.e. 6 consecutive notes separated by a semitone), and vertices separated by perfect $5^{th}$ intervals in sequence.

\begin{table}[h]
    \centering
    \begin{tabular}{|c||c|}
        \hline
        Class of Set & Count \\
        \hline\hline
        $\{ m_i, m_{i+6} \}$ & 6 \\
        \hline
    \end{tabular}
    \vspace{2mm}
    \caption{Maximal independent sets of the 12-vertex minor scale pivot modulation graph.}
    \label{tab:min_ind_set}
\end{table}

\begin{table}[h]
    \centering
    \begin{tabular}{|c||c|}
        \hline
        Class of Set & Count \\
        \hline\hline
        $\{ m_i, m_{i+1}, m_{i+2}, m_{i+3}, m_{i+4}, m_{i+5} \}$ & 12 \\
        \hline
        $\{ m_i, m_{i+1}, m_{i+2}, m_{i+4}, m_{i+5}, m_{i+9} \}$ & 12 \\
        \hline
        $\{ m_i, m_{i+1}, m_{i+2}, m_{i+3}, m_{i+5}, m_{i+10} \}$ & 12 \\
        \hline
        $\{ m_i, m_{i+1}, m_{i+2}, m_{i+5}, m_{i+9}, m_{i+10} \}$ & 12 \\
        \hline
        $\{ m_i, m_{i+2}, m_{i+4}, m_{i+7}, m_{i+9}, m_{i+11} \}$ & 12 \\
        \hline
        $\{ m_i, m_{i+3}, m_{i+4}, m_{i+7}, m_{i+8}, m_{i+11} \}$ & 4 \\
        \hline
    \end{tabular}
    \vspace{2mm}
    \caption{Maximal cliques of the 12-vertex minor scale pivot modulation graph.}
    \label{tab:min_clique}
\end{table}

We describe the group of automorphisms in terms of its generators: 
\begin{eqnarray*} 
& & (m_0, m_1)(m_6, m_7), (m_1, m_2)(m_7, m_8), (m_2, m_3)(m_8, m_9), \\
 & & (m_3, m_4)(m_9, m_{10}), (m_4, m_5)(m_{10}, m_{11}), (m_5, m_{11}).
 \end{eqnarray*} 
This automorphism group has order 46080, with the order of each generator being 2.  

%
%

\subsection{Both Types of Scales}

\begin{figure}[h]
    \centering
    \includegraphics[scale=0.5]{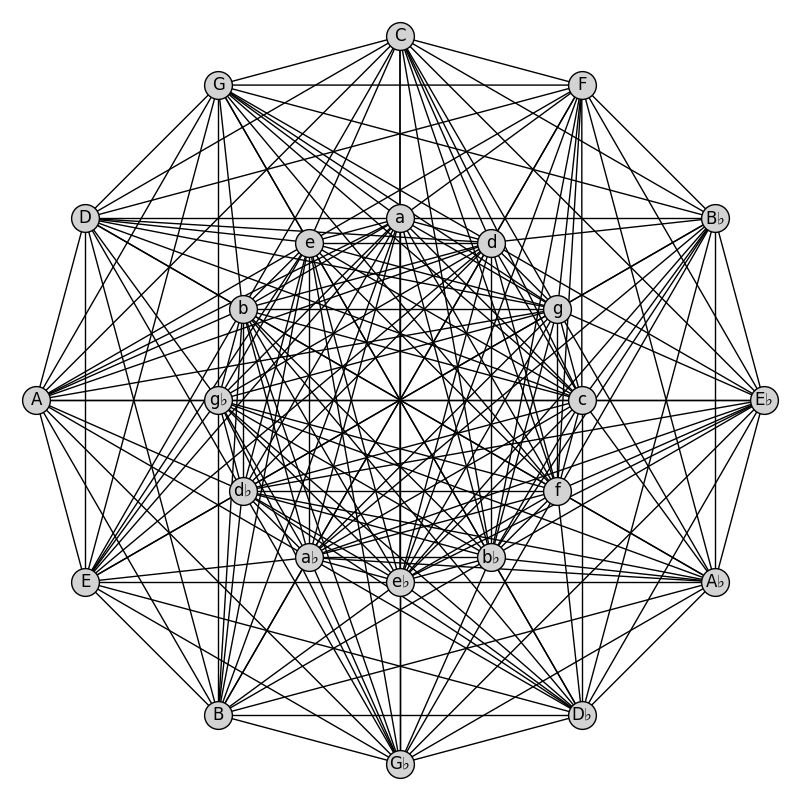}
    \caption{The 24-vertex pivot modulation graph containing both the major-type vertices (outer ring) and minor-type vertices (inner ring).}
    \label{fig:full_graph}
\end{figure}

Thus far we have only considered graphs of pivot modulation involving strictly major or minor flavours of scales.  To not consider a case where modulation between flavours is possible would be to ignore much of the creative realm when it comes to modulation.  Therefore, in this section we examine the 24-vertex graph containing both the major and minor type vertices used previously.  To distinguish between the two types of vertices in the graph, as before we will use upper-case letters to denote tonics of major scale vertices and lower-case letters for the tonics of minor scale vertices (see Figure~\ref{fig:full_graph}).

The outside ring of the graph contains exactly the major-type vertices, and thus the subgraph induced by the outer ring is the pivot modulation graph explored in Section $3.1$.  Similarly, the inside ring contains exactly the minor-type vertices, and thus its induced subgraph is the graph explored in Section $3.2$.  It is the edges between these rings that gives a new and non-trivial pivot modulation graph.  As has been the case thus far, this graph has diameter 2.  It shares with the 12-vertex major pivot modulation graph independence number 3, while uniquely bearing clique number 10.

This graph has 7 classes of maximal independent sets, with all but 1 containing only 2 vertices.  We are now able to see independent sets containing both major and minor type vertices.  Pairs of vertices separated by a tritone appear as maximal independent sets, both being major, minor, or 1 of each.  A pair of major type vertices separated by a semitone forms a maximal independent set, as does a vertex of each type with the minor vertex a single semitone or 8 semitones higher.  The only class of maximal independent set containing 3 vertices is composed of major type vertices that form an augmented triad.  There are 6 classes of maximal cliques in this graph, 2 of which contain only minor type vertices.
Also of interest is the class of maximum cliques: four vertices in each set are major type, and are separated by consecutive perfect $5^{th}$s.  The remaining 6 vertices are minor type, and similarly are separated by consecutive perfect $5^{th}$ intervals.  Note that splitting these sets into strictly their major or minor vertices forms a class of maximal set in their appropriate 12-vertex graph.

\begin{table}[h]
    \centering
    \begin{tabular}{|c||c|}
        \hline
        Class of Set & Count \\
        \hline\hline
        $\{ M_i, M_{i+1} \}$ & 12 \\
        \hline
        $\{ M_i, M_{i+6} \}$ & 6 \\
        \hline
        $\{ M_i, m_{i+1} \}$ & 12 \\
        \hline
        $\{ M_i, m_{i+6} \}$ & 12 \\
        \hline
        $\{ M_i, m_{i+8} \}$ & 12 \\
        \hline
        $\{ m_i, m_{i+6} \}$ & 6 \\
        \hline
        $\{ M_i, M_{i+4}, M_{i+8} \}$ & 4 \\
        \hline
    \end{tabular}
    \vspace{2mm}
    \caption{Maximal independent sets of the 24-vertex pivot modulation graph combining major and minor type vertices.}
    \label{tab:both_ind_set}
\end{table}

\begin{table}[h]
    \centering
    \begin{tabular}{|c||c|}
        \hline
        Class of Set & Count \\
        \hline\hline
        $\{ m_i, m_{i+1}, m_{i+2}, m_{i+3}, m_{i+4}, m_{i+5} \}$ & 12 \\
        \hline
        $\{ m_i, m_{i+1}, m_{i+4}, m_{i+5}, m_{i+8}, m_{i+9} \}$ & 4 \\
        \hline
        $\{ M_i, m_i, m_{i+2}, m_{i+3}, m_{i+4}, m_{i+7}, m_{i+11} \}$ & 12 \\
        \hline
        $\{ M_i, m_i, m_{i+2}, m_{i+3}, m_{i+7}, m_{i+10}, m_{i+11} \}$ & 12 \\
        \hline
        $\{ M_i, M_{i+5}, m_i, m_{i+2}, m_{i+3}, m_{i+4}, m_{i+5}, m_{i+7} \}$ & 12 \\
        \hline
        $\{ M_i, M_{i+2}, M_{i+7}, M_{i+9}, m_i, m_{i+2}, m_{i+4}, m_{i+7}, m_{i+9}, m_{i+11} \}$ & 12 \\
        \hline
    \end{tabular}
    \vspace{2mm}
    \caption{Maximal cliques of the 24-vertex pivot modulation graph combining major and minor type vertices.}
    \label{tab:both_clique}
\end{table}

The automorphism group is isomorphic to $D_{12}$, with the explicit generators: 
\[(M_i, M_{i+1}, ..., M_{i+11})(m_i, m_{i+1}, ..., m_{i+11})\]
and 
\begin{eqnarray*}
&& (M_1, M_{11})(M_2, M_{10})(M_3, M_9) \ldots (M_4, M_8)(M_5, M_7) \cdot \\
&& (m_0, m_2)(m_3, m_{11})(m_4, m_{10})(m_5, m_9)(m_6, m_8).
\end{eqnarray*}

\section{Modulation in The Beatles' Music}

The Beatles hold an impressive discography across a decade of active songwriting.  A body of work has formed for studying their music, or using it in catalogues for data research (for example, \cite{chai, nobile, whissell, hu}).  
In this section we will use the music of The Beatles to connect our theoretical pivot modulation graphs to what is done in practice.

We begin our study of modulation in The Beatles' music with an analysis of 183 songs across their discography.  This list contains the songs from the \emph{isophonics} \footnote{http://isophonics.net/content/reference-annotations-beatles} musical annotation project, and other singles.  Each song was examined for modulations of key according to the framework outlined in Section 3.  This involved following changes in chords for each song, obtained through the \emph{isophonics} project or with cross-reference between \emph{Ultimate Guitar Tabs} and \emph{The Beatles: Complete Scores} \cite{ultimate, beatlescomplete}.  The key(s) present in a song were determined manually through musical analysis.  
All modes of modulation (direct, transitional or pivot) were consdiered, as we wished to find what was actually done in practice.

Out of our list of 183 songs, 77 were found to have at least one modulation.  If a modulation appeared more than once in a song, it was only recorded a single time; 
these modulations would not be unique songwriting decisions, but rather a single decision that is repeated to fit the conventions of song structure.  Once the modulations of a song were found, pivot chords for each modulation were considered.  
If more than a single chord could be a pivot, then that which felt most like the pivot was chosen.  If there was no possibility for a pivot chord, i.e. the modulation was direct or transitional, then it was noted that no pivot chord was used.  Chords that extend beyond triads (i.e. $7^{th}$, $9^{th}$, etc. chords) were treated as the fundamental triad of their structure.
For example, the modulations in ``Think For Yourself'' by George Harrison can be expressed by $A\text{mi} \xrightarrow{} G \xrightarrow{} A\text{mi}$; the verses of this song are in the key of $A$mi, while the choruses are in $G$.  Thus we have modulation from $A$mi to $G$ and from $G$ to $A$mi.  In the former a pivot triad of $C$maj is used, while in the latter the pivot is $A$mi.

\begin{figure}[h]
    \centering
    \includegraphics[scale=0.55]{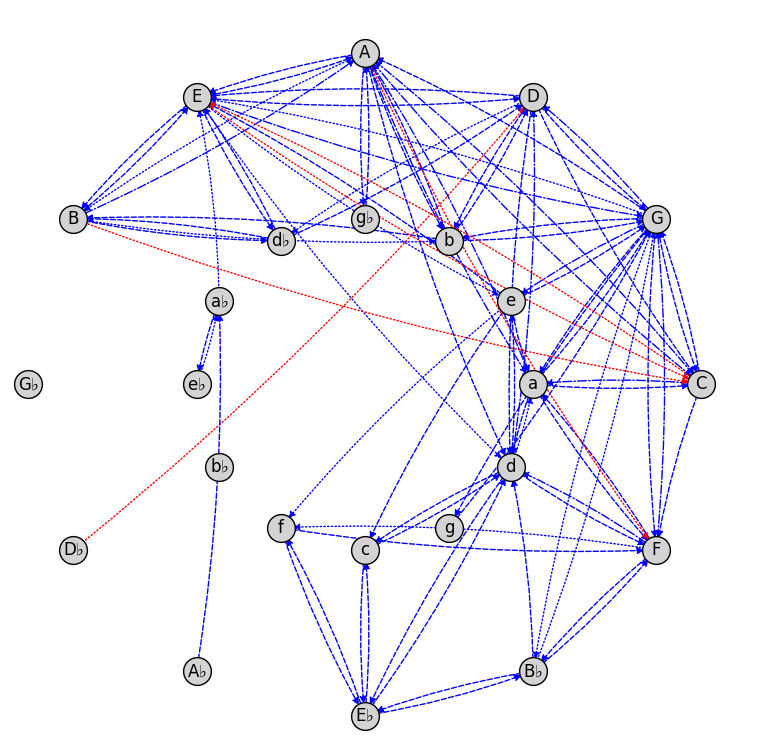}
    \caption{Modulations found in the music of The Beatles.  Blue edges indicate modulations allowed by a pivot, while red indicates otherwise.  Some modulations were only performed with pivots (dashed), some only by directly modulating (dotted), or both ways (dotted and dashed).}
    \label{fig:beatles_edges}
\end{figure}

All of the modulations found in The Beatles' songs can be seen in the graph in Figure \ref{fig:beatles_edges}.  We will use the vertex arrangement from the full pivot modulation graph in Section 3.2 to show what modulations they use.  As there was a priori no guarantee that all the modulations observed were reversed, the graph formed is directed.  Edges that are blue represent those permitted by pivot modulation, and red represents otherwise (only possible through direct modulation).  A dashed edge indicates a modulation only occurred through use of a pivot, a dotted edge indicates a modulation only occurred directly and without a pivot, while a dotted and dashed edge marks a modulation that occurred both with and without a pivot.  There are in total 89 directed edges in the graph, with 59 being strictly pivot modulation, 18 strictly direct modulation, and 12 being of both mechanisms.

\noindent Only 4 modulations across three songs are not permitted by a pivot, hence non-pivoting modulations were rare occurrences in The Beatles' canon.  Interestingly, there are two vertices not involved in any modulation: $G\flat$ and $b\flat$.  In contrast, the most active vertex is $G$, having in-degree 9 and out-degree 10.

\begin{figure}[h]
    \centering
    \includegraphics[scale=0.5]{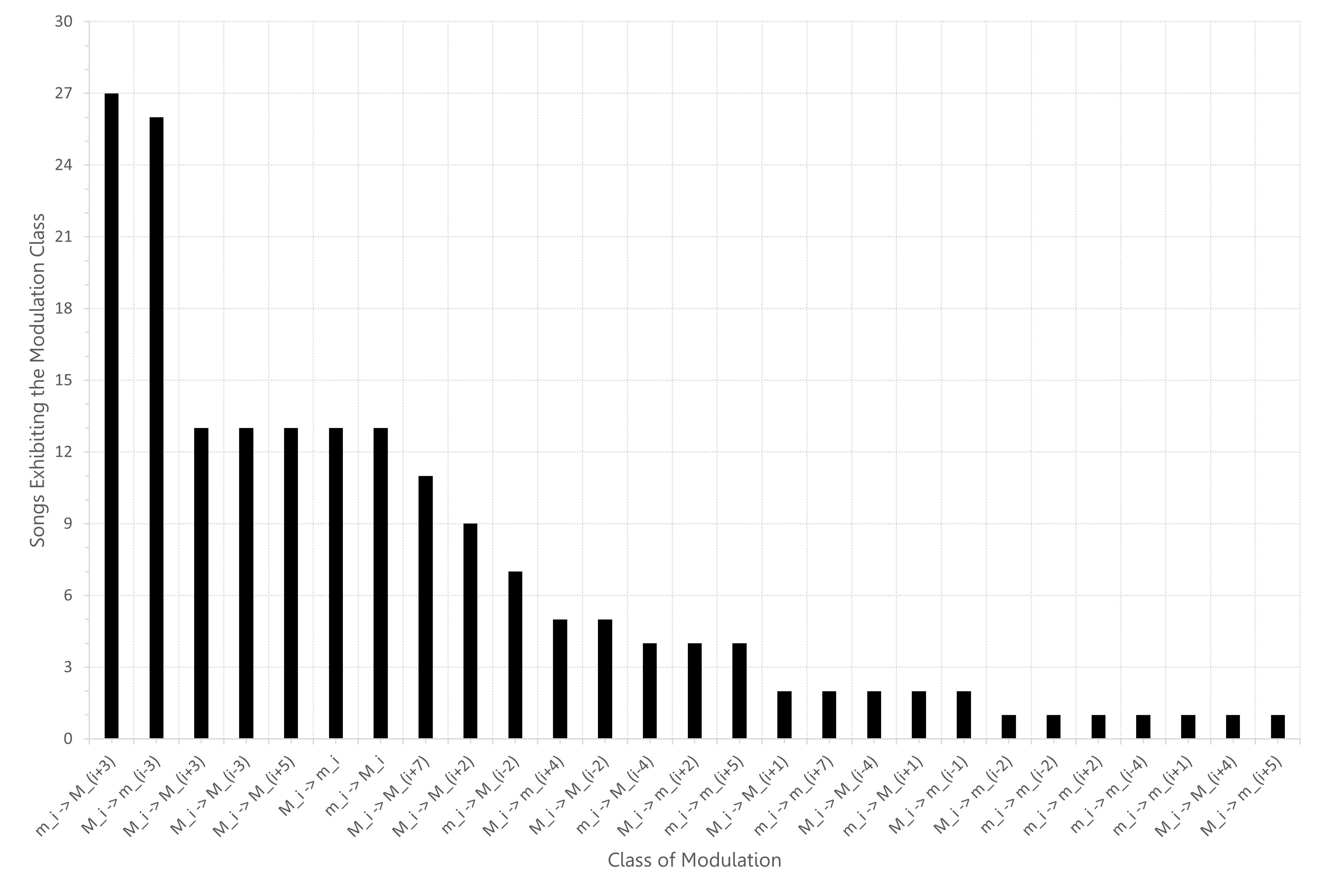}
    \caption{Counting the number of song appearances for each modulation class present in The Beatles' music.}
    \label{fig:mod_counts}
\end{figure}

We are also interested in modulations of a certain class.  
Intuitively, modulations such as $A$mi $\xrightarrow{} C$ and $E$mi $\xrightarrow{} G$ should be of the same class, because both initial and destination keys are the same flavour of scale and the semitone distances between the initial keys equals that of the destination keys.  Given a modulation $x_i \xrightarrow{} y_j$ where $x, y \in \{M, m\}$ and $i, j \in \mathbb{Z}_{12}$, the set of transposed modulations $x_{i+k} \xrightarrow{} y_{j+k}$ ($k \in {\mathbb Z}_12$) forms its equivalence class.  
The example modulations above belong to the $m_i \xrightarrow{} M_{i+3}$ class.  Across the 77 songs found to have a modulation, there are 27 unique classes of modulation (a testament to the harmonic creativity of John, Paul and George).  Three of these classes do not have pivot chords: $M_i \xrightarrow{} M_{i+1}$, $M_i \xrightarrow{} M_{i+4}$, $M_i \xrightarrow{} M_{i-4}$.  We also make the following observations: 7 classes are from a major vertex to a minor vertex, 5 are a minor vertex to a major vertex, 9 are between major vertices, and 6 are between minor vertices.  To see the modulation tendencies of The Beatles, we counted the number of modulations present in each class from the 77 songs containing modulation.  The largest modulation classes are $m_i \xrightarrow{} M_{i+3}$ and $M_i \xrightarrow{} m_{i-3}$, having 27 and 26 instances of modulation, respectively.  These classes are simply the common modulations between relative major and minor type scales in each direction.  The second most counted modulation classes are $M_i \xrightarrow{} m_i$, $m_i \xrightarrow{} M_i$ (parallel pairs), $M_i \xrightarrow{} M_{i+3}$, $M_i \xrightarrow{} M_{i-3}$, and $M_i \xrightarrow{} M_{i+5}$, all with 13 instances.  The reverse modulation of the final class, $M_i \xrightarrow{} M_{i+7}$, has a similar count of 11. Figure~\ref{fig:mod_counts} summarizes the different classes of modulations in The Beatles' songs.

\begin{figure}[h]
    \centering
    \subfigure[Graph depicting modulation for ``Doctor Robert'' and ``Blackbird''.]{\includegraphics[scale=0.3]{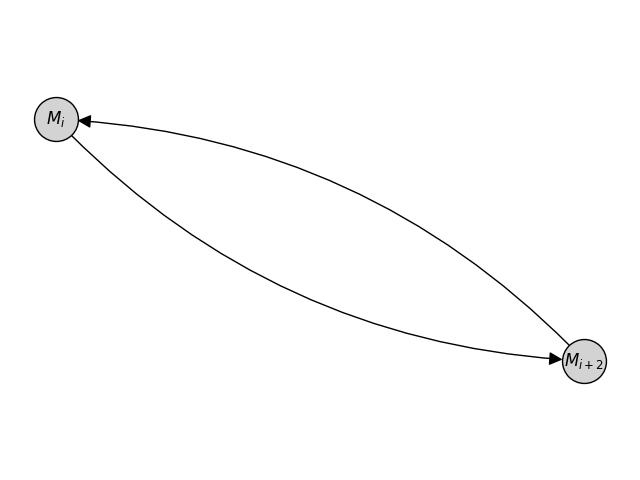}}
    \hspace{30pt}
    \subfigure[Graph depicting modulation for ``Good Day Sunshine''.]{\includegraphics[scale=0.3]{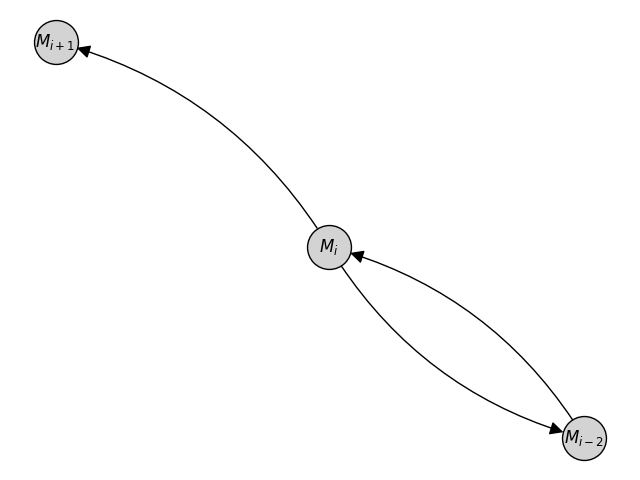}}
    \caption{Two examples of directed graphs that minimally describe the modulations of Beatles songs.}
    \label{fig:example_subgraph}
\end{figure}

Considering classes of modulation can allow us to identify structures of modulation present in The Beatles' music.  Using modulation between vertices of the form $M_i$ or $m_i$, we can form a small directed graph that entirely describes the modulations of a song.  Let us illustrate this with an example.  The song ``Doctor Robert'' contains modulations from $A \xrightarrow{} B$ and $B \xrightarrow{} A$.  More generally, these modulations are $M_i \xrightarrow{} M_{i+2}$ and $M_{i+2} \xrightarrow{} M_i$.  However, ``Blackbird'' contains modulations $G \xrightarrow{} F$ and $F \xrightarrow{} G$, which are described in the same way.  Therefore the directed cycle of vertices $M_i$ and $M_{i+2}$ describes the modulation activity of both of these songs.  The graph which describes the modulations of ``Good Day Sunshine'' contains the directed cycle between $M_i$ and $M_{i+2}$, but it is a proper containment, as there is a different type of modulation as well (see Figure~\ref{fig:example_subgraph}).  
There are 35 unique directed graphs across all songs, that illustrate the variety of modulations in the songs.

\section{Discussion}


The discography of The Beatles provided a sizable catalogue of songs whose modulations could be analyzed through our modulation framework.  We were able to capture the majority of modulations present in these songs with our rules of pivot modulation.  Noted in the spreadsheet of song analyses is the occasional use of the Neapolitan $6^{th}$, or the tritone substitution, as a method for modulation which may not be captured by our framework.  This is perhaps an influence of classical music and jazz in the music of The Beatles.

An important factor to our Beatles analysis is human determination of song key and modulations.  Our judgement of song key(s) were based on chord transcriptions performed by hand \cite{harte, beatlescomplete, ultimate}, and our own knowledge of music.  Of course, such practices may lend themselves to human error.  Methods of key or modulation detection by computational and machine learning techniques \cite{feisthauer, korzeniowski, lopez, mearns} may be favoured in further graph-theoretic analysis of modulation, especially on very large data sets for the purposes of time efficiency and error reduction.

Future work with graphs of modulation may include examining graphs resulting from different treatments of modulation (such as mentioned above), relating them to other genres of music.  Other song catalogues spanning different genres or artists may be of interest, especially to make genre-to-genre comparisons of modulation.  We hope that the ideas and results of this paper are of interest to both mathematicians and musicians.  Perhaps modulation graphs could be a creative tool for composition and may direct the songwriting choices of composers in ways previously unseen.

\section*{Acknowledgements}  
 
J. Brown acknowledges research support from the Natural Sciences and Engineering Research Council of Canada (NSERC), grant RGPIN 2018-05227.


\bibliographystyle{tfs}
\bibliography{mod_bib}

\end{document}